\documentclass{elsart}

\begin{document}
\begin{frontmatter}
\hfill{}Solid State Communications, {\bf 97}, \#12 (1996)

\title{New Mechanism of Magnetoresistance in Bulk Semiconductors: 
Boundary Condition Effects}

\author{G. Gonzalez de la Cruz,\and Yu.\ G. Gurevich}
\address{Departamento de F\'{\i}sica, 
Centro de Investigaci\'{o}n y 
Estudios Avanzados del Instituto Polit\'{e}cnico Nacional, 
Apartado Postal 14--740, 07000 M\'{e}xico D.F.}
\author{V. V. Prosentsov}
\address{Deparment of Physics, 
Politechnical University, 
Kharkov 310002, Ukraine.}

\begin{abstract}
We consider the electronic transport in bounded semiconductors  in the
presence of an external magnetic field. Taking into account appropriate 
boundary conditions for the current density at the contacts, a change 
in the magnetoresistance of bulk semiconductors is found as compared 
with the usual theory of galvanomagnetic effects in boundless media. 
New mechanism in magnetoresistance connected with the boundary 
conditions arises. In particular, even when the relaxation time is 
independent of the electron energy, magnetoresistance is not vanish.
\end{abstract}
\end{frontmatter}

It is well known that magnetoresistance means the increasing in 
the electrical resistance of a metal or semiconductor when placed in 
a magnetic field. The effect of greatest interest is the transverse 
magnetoresistance, which is usually studied in the following 
geometrical arrangement: a long conducting channel is directed along 
the $x$-axis, and a uniform electric field $E_x$  is established in 
the channel by means of an external power supply. A uniform magnetic 
field  $B$ is applied along the $y$-axis, thus normal to the axis of 
the channel. As a result of the geometry of the sample the carriers 
are  deviated from the $x$ direction and an electric field (Hall 
field) appears as a consequence of the accumulation of carriers on 
the surface of the sample in the $z$ direction. The electrical 
current along the Hall field is usually considered zero in the 
standard theory of magnetoresistance after reach steady state 
conditions.

In the geometry described before, the effect of weak magnetic field 
$\omega_H \tau\ll 1$, where $\omega_H$ is the cyclotron frequency 
and $\tau$ the relaxation time, is to increase the resistance by an 
additive term proportional to $B^2$.

There are two important mechanisms in magnetoresistance in the 
conventional theory: one of them is connected with the dependence of 
the electric conductivity on the magnetic field (physical 
magnetoresistance), and the other one has to do with the absence of 
the Hall electric current (geometrical 
magnetoresistance) \cite{Madelung}. In the case when the relaxation 
time is independent of electron energy, the physical and geometrical 
contributions to the magnetoresistance cancel each other and the 
total magnetoresistance is zero \cite{BBG}.

In standard theory, magnetoresistance in isotropic semiconductors has 
only significant values in the case when the relaxation time depends 
on the energy, because the Lorentz force acts in a way on carriers 
with different velocities.

In all the theoretical papers, as far as galvanomagnetic effects in 
bulk semiconductors are concerned, the electric field $E_x$ is 
usually considered independent of the magnetic field and coordinates 
as well as the size of the sample.  However, in reality it is usual 
to establish some specific boundary conditions on the surfaces of the 
sample and as a consequence, in general, the electrostatic field is 
function of the coordinates and magnetic field in this bounded 
semiconductor (the corresponding result is written below).

Most of the boundary conditions used in the investigation of the 
galvanomagnetic effects in semiconductors have been focused only by 
considering  a finite system along the Hall field \cite{BBG}. Very 
recently Akhiezer, Gurevich and Zakirov \cite{AGZ}, developed a 
theory of Hall effect and magnetoresistance in weak electric and 
magnetic fields for semiconductor samples of finite dimensions in all 
directions. The existence of vortex  electric currents as a 
consequence of the finiteness of the sample, and  magnetoresistance 
were obtained as a linear function of the magnetic field. The current 
distribution  in this bounded system was computed and compared for 
different sample geometries. However, in the limit when the dimensions 
of the semiconductor are very large, the effects of the magnetic field 
disappears.

Sokolov et.al. \cite{SG}, have been discussed that in a sample with 
arbitrary geometry in weak electric and magnetic fields the expression 
of the magnetoresistance, proportional to ${(\omega_H\tau)}^2$, changes 
substantially. However, when the dimensions of the specimen are very 
large, the usual result is recovered.

In this letter, we will quantitatively extend the analysis of 
magnetoresistance in bounded semiconductors in all directions. The 
analysis will show that it should be possible to observe a new and 
rather interesting electronic transport due to the dependence of the 
electric field on the magnetic field. In particular, specific 
definition of the boundary conditions for the potential at the 
surfaces, leads a change of  the magnetoresistance in the limit of 
infinite system (bulk semiconductors) and it is not vanish when the 
relaxation time is independent of energy. This remarkable result comes 
from the dependence of the $x$ component of the electric field on the 
magnetic field by a term proportional to ${(\omega_H\tau)}^2$ which is 
finite when the dimensions of the sample are very large. This unusual 
behavior  alter significantly the usual theory of galvanomagnetic 
effects developed for boundless media and conducting channels.  

We shall assume that a semiconductor has the shape of a parallelepiped 
bounded by the $x=\pm a$, $y=\pm c$ and $z=\pm b$ planes and the 
$x=\pm a$ planes have the current contacts, while the magnetic field 
is directed along the $y$-axis. In this problem the geometry clearly 
is two-dimensional (all the quantities only depend of $x$ and $z$). 
The electrostatic potential distribution in the sample can be found as 
a function of the coordinates and magnetic field by considering the 
current continuity equation. Under steady-state conditions, the 
electric current density ${\bf j}$ is described by the following 
expression \cite{GM}:
\begin{equation}
{\bf j} = 
e^2 I_{10} {\bf E}  + \left(e^2 I_{20} {\bf E}\right) \times {\bf h}
\label{eq:1}
\end{equation}
and  the continuity equation is:
\begin{equation}
\nabla\cdot{\bf j} = 0.
\label{eq:2}
\end{equation}

Assuming that the relaxation time is independent of electron energy 
(in this case magnetoresistance becomes zero in the conventional 
theory) and that the electron gas is nondegenerate (Maxwell statistics), 
the electron density with a quadratic and isotropic energy-momentum 
relation $\varepsilon=p^{2}/ 2m$ is
\[
n = \frac{1}{2}{\left(\frac{2m}{\pi\hbar^2}\right)}^{3/2}
T^{3/2}\exp\left(\mu/T_{e}\right)
\]
the electric field is given by ${\bf E}= - \nabla\varphi$, 
$\varphi$ is the electrostatical potential, 
$\mu$ the chemical potential, $T$ the temperature, $\hbar={\bf B}/B$ and
\begin{equation}
I_{10} = \frac{n\tau}{m} \left[ 1 - (\omega_{H}\tau)^{2}\right],\qquad
I_{20} = \frac{n\tau}{m} \omega_{H}\tau .
\label{eq:Iint}
\end{equation}

The expression for the average value of the current density over the 
cross section of the semiconductor significant for the 
magnetoresistance is given by
\begin{equation}
\bar{j_x} = \frac{1}{2b}\int_{-b}^{+b} j_x(x,z) \, dz .
\label{eq:5}
\end{equation}
The expression for $j_{x}(x,z)$ can be obtained from Eq.~(1) and (2) 
with appropriate boundary conditions for the current density at the  
contacts i.e.
\begin{equation}
\left. \varphi\right|_{x = \pm a} = \pm \frac{\Delta\varphi}{2} 
\end{equation}
where $\Delta\varphi$ is the difference potential applied to the sample. 
In addition, it is natural to choose $j_z = 0$ at $z = \pm b$, no 
contacts along the Hall field. 

Substituting Eq.~(3) into Eq.~(1) and (2) and assuming that the electric 
field is weak, so that nonlinear effects (Joule heating) are negligible 
in Eq.~(\ref{eq:2}), we can write the equations for the potential 
$\varphi(x,z)$ in the sample up to terms proportional to $B^{2}$ as
\begin{equation}
\nabla^2 \varphi(x,z) = 0,
\label{eq:7}
\end{equation}

The potential distribution satisfies the following equations at the 
surface of the sample where $\left. j_{z} \right|_{z=\pm b} = 0$:
\begin{equation}
\left.
\frac{\partial\varphi}{\partial z} 
 + \omega_{H}\tau_{0} \frac{\partial\varphi}{\partial x}
\right|_{z=\pm b} =0.
\end{equation}

Assuming $\Delta\varphi$ to be small, we naturally seek solution of 
the Eq.~(6) in the form
\begin{equation}
\varphi = \varphi_{0}\Delta\varphi 
	+ \varphi_{1}\Delta\varphi(\omega_{H}\tau_{0})
	+ \varphi_{2}(\omega_{H}\tau_{0})^{2}
\end{equation}

The quadratic correction to the potential with respect to the magnetic 
field is the important contribution to the magnetoresistance. 
Substituting Eq.~(8) into Eq.~(5) and (6), for the linear approximation 
in $\Delta\varphi$ and zeroth approximation with respect to the magnetic 
field, we find the well known expressions for the potential, the 
electric field and the current density:
\begin{equation}
\varphi_{0} = - \frac{1}{2a}x,\quad
E_{x}^{0} = \frac{\Delta\varphi}{2a},\quad
E_{z}^{0} = 0 ;\qquad
j_{x}^{0} = \sigma_{0} \frac{\Delta\varphi}{2a};\quad
j_{z}^{0} = 0,
\end{equation}
where
\[
\sigma_{0} = \frac{n e^{2}\tau}{m}
\]
is the electric conductivity of electrons in the absence of magnetic 
field.

For linear approximation in the magnetic field, 
$\Delta\varphi(\omega_{H}\tau)$, we also employ Eq.~(6) and (8), 
we can easily obtaine the following differential equation 
for $\varphi_1$
\begin{equation}
\nabla^{2}\varphi_{1}(x,z)= 0
\end{equation}
indeed, with the conditions (5) and (7) satisfied, we may write
\begin{equation}
\left.
\varphi_{1}
\right|_{x=\pm a} =0,\quad
\left.
\frac{\partial\varphi_1}{\partial z}
\right|_{z=\pm b}= \frac{1}{2a}.
\end{equation}

Combining Eq.~(10) and (11), we find the correction for the potential 
distribution to first order approximation with respect to the magnetic 
field
\begin{equation}
\varphi_{1} = \frac{1}{a^{2}} 
	\sum_{n=0}^{\infty} \frac{(-1)^{n}}{\alpha_{n}^{2}} 
	\frac{\sinh{\alpha_{n}z}}{\cosh{\alpha_{n}b}} 
	\cos{\alpha_{n}x} 
.
\end{equation}
Here
\[
\alpha_{n} = \frac{\pi}{2a}(2n + 1).
\]

It follows from Eq.~(1) and Eq.~(12), the $x$-component of the current 
dencity to first order in the magnetic field is given by
\begin{equation}
j_{x} = - \sigma_{0} \frac{\partial\varphi_{0}}{\partial x} 
	\Delta\varphi
	-\sigma_{0} \frac{\partial\varphi_{1}}{\partial x}
	\Delta\varphi(\omega_{H}\tau) 
=
j_{0}\Delta\varphi + j_{1} \Delta\varphi(\omega_{H}\tau)
\end{equation}
with
\begin{equation}
j_{1} = \frac{\sigma_{0}}{a^{2}} 
	\sum_{n=0}^{\infty}\frac{(-1)^{n}}{\alpha_{n}} 
	\frac{\sinh{\alpha_{n}z}}{\cosh{\alpha_{n}b}} 
	\sinh{\alpha_{n}x}
\end{equation}
It can see from expression (4) that the average value of the 
contribution to the current density in the linear approximation 
vanishes i.e. there is not magnetoresistance proportional to $B$, 
in other words
\[
\bar{j_{1}}(x) = \frac{1}{2b}\int_{-b}^{+b}j_{1}(x,z)\, dz = 0.
\]

Moving on to the calculations of the coefficient of $\varphi_{2}(x,z)$, 
we begin with the explicit equation which determine this quantity: it 
satisfies a similar expression that Eq.~(10) and from the boundary 
conditions Eq.~(5) and (7) we can write
\begin{equation}
\left.
\varphi_{2}
\right|_{x=\pm a} =0, \qquad
\left.
\frac{\varphi_{2}}{\partial z} + \frac{\partial\varphi_{1}}{\partial x}
\right|_{z=\pm b}
=0 .
\end{equation}
In order to evaluate $\varphi_{2}$ it is convenient to use the 
following Fourier expansion of $\sin{\alpha_{n}x}$ in expression (12) 
only valid inside the semiconductor
\[
\sin{\alpha_{n}x} = 
(-1)^{n} \frac{2}{a} \sum_{m=0}^{\infty} (-1)^{m} 
\frac{\beta_{m}}{\beta_{m}^{2} - \alpha_{n}^{2}} \sin{\beta_{m}x}, 
\quad
\beta_{m} = \frac{m\pi}{a} 
.
\]

After some tedious but straightforward algebra, the general expression 
for $\varphi_{2}$ is given by
\[
\varphi_{2} (x,z) = 
\frac{2}{a^{3}} 
\sum_{n,m=0}^{\infty} (-1)^{m+1} 
\frac{\tanh{\alpha_{n}b}}{\alpha_{n}}
\frac{\cosh{\beta_{m}z}}{\sinh{\beta_{m}z}}
\frac{sinh{\beta_{m}x}}{\beta_{m}^{2} - \alpha_{m}^{2}}.
\]
Now, it is possible to write the $x$-component to the current density in 
second order to the magnatic field
\[
j_{x} = 
j_{0}\Delta\varphi 
+ j_{1}\Delta\varphi(\omega_{H}\tau) 
+ j_{2} \Delta\varphi(\omega_{H}\tau)^{2} 
,
\]
where
\begin{equation}
j_{2} = \sigma_{0} 
\left(
\frac{\partial\varphi_{0}}{\partial x} 
+ \frac{\partial\varphi_{1}}{\partial z} 
- \frac{\partial\varphi_{2}}{\partial x}
\right) 
.
\end{equation}
It is important to note that in the latter expression in second member, 
the first and second terms represent the physical and geometrical 
contribution to the magnetoresistance respectivly. However, the last 
term is a new expression conected with the dependence of the 
$x$-component of the electric field in second order on the 
magnetic field.

Then the average value of the contribution to the current density in 
the second order approximatin is given by
\begin{equation}
\bar j_{2} = \sigma_{0} 
\left(
- \frac{1}{2a} 
- \frac{1}{a^{3}b}
\sum_{n=0}^{\infty} 
\frac{\tanh{\alpha_{n}b}}{\alpha_{n}^{3}}
\right)
\end{equation}
If $a\gg b$ (bulk semiconductor), from Eq.~(17), we can obtaine the 
following expression
\[
\bar{j_{2}} = - \frac{\sigma_{0}}{a}
\]
It is straightforward to show, that in this limit the sum of the 
physical and geometrical contribution in Eq.~(16) is vanish, this 
corresponds to the result of the conventional theory of 
magnetoresistance. Neverless, the extr term in Eq.~(16) corresponds 
to the contribution to the $x$ component of the electric field in 
second order to the magnatic field and this is not vanish in this limit.

The reason why the magnetoresistance is not zero in this theory comes 
from the behaviour of the potential near the contacts, in the regions 
$a -|x|\ll b$. In this region $\varphi_{1} (x, z=b)$ changes from $0$ 
at the contacts, see Eq.~(11), to $b/2a$, see Eq. ~(12), in the limit 
$a\gg b$. This means that in the boundary conditions for $\varphi_{2}$, 
Eq.~(15), $\partial\varphi_{1}/\partial x \ge 1/a$.

For the case when the relaxation time depends on the energy we obtain 
besides the usual results of magnetoresistance theory a different 
expression of the new term. It is worth to mention that in  this 
situation the Ettingshausen effect (see Ref.~\cite{BBG}) plays an 
important role in magnetoresistance.

\ack
This work is sponsored in part by Consejo Nacional de Ciencia y 
Tecnologia (CONACYT), Mexico.

\end{document}